# Novel Method to Estimate Kinetic Microparameters from Dynamic Whole-Body Imaging in Regular-Axial Field-of-View PET Scanners

Kyung-Nam Lee[1,2], Arman Rahmim[1,2,3], Carlos Uribe[1,3,4]

[1]Department of Integrative Oncology, BC Cancer Research Institute, Vancouver, Canada

[2]Department of Physics & Astronomy, University of British Columbia, Vancouver, Canada

[3]Department of Radiology, University of British Columbia, Vancouver, Canada

[4]Functional Imaging, BC Cancer, Vancouver, Canada

## Abstract

**Purpose:** For whole-body (WB) kinetic modeling based on a typical PET scanner, a multi-pass multi-bed scanning protocol is necessary given the limited axial field-of-view. Such a protocol introduces loss of early-dynamics in time-activity curves (TACs) and sparsity in TAC measurements, inducing uncertainty in parameter estimation when using least-squares estimation (LSE) (i.e., common standard) especially for kinetic microparameters. We present a method to reliably estimate microparameters, enabling accurate parametric imaging, on regular-axial field-of-view PET scanners.

**Materials and Methods:** Our method, denoted parameter combination-driven estimation (PCDE), relies on generation of reference truth TAC database, and subsequently selecting, the best parameter combination as the one arriving at TAC with highest total similarity score (TSS), focusing on general image quality, overall visibility, and tumor detectability metrics. Our technique has two distinctive characteristics: 1) improved probability of having one-on-one mapping between early and late dynamics in TACs (the former missing from typical protocols), and 2) use of multiple aspects of TACs in selection of best fits. To evaluate our method against conventional LSE, we plotted trade-off curves for noise and bias. In addition, the overall signal-to-noise ratio (SNR) and spatial noise were calculated and compared. Furthermore, the contrast-to-noise ratio (CNR) and tumor-to-background ratio (TBR) were also calculated. We also tested our proposed method on patient data ([18]F-DCFPyL PSMA PET/CT scans) to further verify clinical applicability.

**Results:** Significantly improved general image quality performance was verified in microparametric images (e.g. noise-bias trade-off performance). The overall visibility and tumor detectability were also improved. Finally, for our patient studies, improved overall visibility and tumor detectability were demonstrated in micoparametric images, compared to use of conventional parameter estimation.

**Conclusion:** The proposed method provides improved microkinetic parametric images compared to common standard in terms of general image quality, overall visibility, and tumor detectability.

# 1. Introduction

Clinical diagnosis and treatment response monitoring of localized and metastatic cancers have benefited remarkably from whole-body (WB) positron emission tomography/computed tomography (PET/CT) imaging[1–7]. Currently, the standardized uptake value (SUV) is the metric used to measure metabolic activity from quantitative images. PET tracer distribution is a dynamic process altered by several factors that vary considerably depending on the organ, region of interest (ROI), patient, and time of scan[1,8]. Hence, static SUV images are time-dependent, which is undesirable for use in quantitative studies. With the additional use of tracer kinetic modeling techniques that require dynamic PET scanning, there is the potential for substantially improving the type and quality of information of the biological and physiological processes in tissue[1,2] that is not time-dependent. This can enable further clinical benefits from PET images through quantitative analysis. Many studies have shown that kinetic compartment modeling can improve both tumor characterization and treatment response monitoring[2,9–13].

Nonetheless, dynamic PET protocols have been confined to a single-bed position, limiting the axial field-of-view of parametric images to ~15-25 [cm], and have not been translated to multi-bed positions (i.e., WB). However, it is more desirable to inspect disseminated diseases and this has been gaining increasing attention[3–7].

To achieve four-dimensional (4D) WB PET acquisition, the following three challenges present themselves: (1) long acquisition times, (2) few dynamic frames at each bed (i.e., sparsity of data), and (3) noninvasive quantification of rapid early kinetics in the plasma. Karakatsanis et al. optimized the scanning protocol through extensive Monte Carlo simulation studies[8,14]. They proposed an optimal protocol for input function estimation and dynamic WB dataset generation, which comprises two sequential scanning steps: (1) an initial 6 min single-bed dynamic scan over the cardiac region to generate an image-derived input function (addressing challenge number 3) and (2) a sequence of six multi-bed multi-pass WB scans to capture the late dynamics of the tracer in the blood plasma and tissue.

Although the optimal protocol allows for WB kinetic modeling, it was optimized for the measurement of macroparameters, specifically the net influx rate from the plasma into the 2nd compartment in the two-tissue compartment model (i.e., $K_i$).; macroparameters are lumped constants comprised of several microparameters. Hence, this method is not the most appropriate protocol for microparameter estimation if least squares estimation (LSE) is exploited.

Two factors can contribute to uncertainty in microparameter estimation: (1) the loss of early dynamics of time activity (i.e., the loss of near-peak data), except for the chest region in the FOV of the first 6 minutes of the acquisition, and (2) sparsity of measured data (i.e., 5-6 min between scans of the same anatomical region). Due to these factors, the estimation of microparameters for WB kinetic modeling has not been fully implemented in cases where a typical PET scanner (i.e. axial FOV between 15-30 cm) is the only available option for dynamic scans. However, the detailed explanatory power of microparameter estimation in assessing the biochemical status of tissues can significantly enhance effectiveness and flexibility in clinical applications, surpassing the capabilities of macroparameters.

We aimed to develop a novel method to enable accurate kinetic modeling including the estimation of the microparameters using multi-pass protocols in typical PET scanner-based WB imaging. We refer to this new method as parameter combination-driven estimation (PCDE). We evaluated the method's in terms of image quality, overall visibility, and tumor detectability compared to LSE (i.e., common standard).

## 2. Methods

### 2.1. Generating Simulated Data

#### 2.1.1. Noise-free Images

To generate ground-truth PET images, we employed the 4D extended cardiac-torso (XCAT) phantom[15], which is well-validated and widely used for performance testing of new algorithms or approaches in numerous areas of medical imaging. The dynamics of the activity distribution assigned to each ROI in the XCAT phantom were based on realistic fluorodeoxyglucose (FDG) kinetic microparameters, as reported in the literature[8,16] and presented in Tables 1 and 2. Volumes of organs are also shown in Table 1 (note that for quantitative analysis, entire ROI volumes equal to organ sizes were used). In this study, the reversible uptake process rate $k_4$ was assumed to be zero.

**Table 1. Ground truths of kinetic microparameters for normal whole-body organs.**

|  | $K_1$ | $k_2$ | $k_3$ |
|---|---|---|---|
| Brain (vol.: 129.8 ml) | 0.13 | 0.63 | 0.19 |
| Thyroid (vol.: 25.1 ml) | 0.97 | 1.00 | 0.07 |
| Myocardium (vol.: 159.8 ml) | 0.82 | 1.00 | 0.19 |
| Spleen (vol.: 170.4 ml) | 0.88 | 1.00 | 0.04 |
| Pancreas (vol. 138.6 ml) | 0.36 | 1.00 | 0.08 |
| Kidney (vol.: 325.2 ml) | 0.70 | 1.00 | 0.18 |
| Liver (vol.: 1767.9 ml) | 0.86 | 0.98 | 0.01 |
| Lung (vol.: 2757.7 ml) | 0.11 | 0.74 | 0.02 |

**Table 2. Ground truths of kinetic macroparameters for tumors.**

|  | $K_1$ | $k_2$ | $k_3$ |
|---|---|---|---|
| Lung | 0.3 | 0.86 | 0.05 |
| Liver | 0.24 | 0.78 | 0.1 |

*Tumor shape and size: sphere with 1.5 cm diameter.

A plasma input function was created based on Feng's model[17], and the basic formula of the two-tissue compartment model (2TCM) was used to calculate true activities over time as follows:

$$C_{PET}(t) = \frac{K_1}{\alpha_2 - \alpha_1}[(k_3 + k_4 - \alpha_1)e^{-\alpha_1 t} + (\alpha_2 - k_3 - k_4)e^{-\alpha_2 t}] \otimes C_p(t) \quad (1)$$

$$\alpha_{1,2} = \frac{k_2 + k_3 + k_4 \mp \sqrt{(k_2 + k_3 + k_4)^2 - 4k_2 k_4}}{2} \quad (2)$$

where $C_{PET}$ and $C_p$ denote the measured PET concentration and plasma concentration input function, respectively, and $\otimes$ is the convolution operator. $K_1$ and $k_2$ are the influx and efflux rate constants between the plasma and first tissue compartments, and $k_3$ and $k_4$ represent the influx and efflux rate constants between the first and second tissue compartments, respectively. In above formulas and present investigation, blood volume is not included, but it can be easily added in future efforts within the framework proposed in this work.

To alleviate the long scan time (i.e., one of the disadvantages of dynamic acquisition), we limited the total acquisition duration to 40 min after injection. We also only used the data between 10-40 min post-injection (PI) to simulate the loss of early dynamics due to first-phase scanning of the cardiac region. Based on true kinetic parameters (i.e., Tables 1 and 2) and the predefined scanning protocol for virtual dynamic set (i.e., Table 3), the calculated concentrations with time were assigned for each ROI in the XCAT input files to generate noise-free XCAT phantom images.

**Table 3. Scanning protocol for virtual dynamic dataset.**

| Item | Value |
|---|---|
| Total acquisition time (cardiac + WB) | 40 min |
| Image acquisition for WB | *10-40 min |
| Time interval | 5 min |
| # of passes | 7 |
| # of beds | 5 |

*10 min was assumed to simulate a scenario worse than that of the protocol proposed by Karakatsanis. Time: Post-injection time.

2.1.2. Noise Realizations

To add realistic noise, we employed a Dynamic PET Simulator of Tracers via Emission Projection[18,19] (dPETSTEP), which is a fast and simple tool to simulate dynamic PET as an alternative to Monte Carlo simulation. Noise-free XCAT phantom images and attenuation maps were used as input data to generate a realistic (i.e., noisy) dynamic PET dataset. The validated settings for the

GE Discovery LS scanner[18] were used with the ordered subset expectation maximization (OSEM) algorithm. Table 4 summarizes the reconstruction settings for $d_{PETSTEP}$.

**Table 4. Summary of reconstruction settings.**

| Item | Value |
| --- | --- |
| Radial bins | 283 |
| Projection angles | 336 |
| OSEM iterations | 1-5 |
| OSEM subsets | 24 |
| PSF | 5.1 mm |
| Post-filter XY | 6 mm Gaussian |
| Post-filter Z | [1 2 1]/4 |
| *Reconstructed matrix per bed | 165 x 165 x 35 |
| Reconstructed voxel size | 2 x 2 x 4.25 mm$^3$ |
| Noise realizations | 10 |

*Reconstructed matrix for the entire body: 165 × 165 × 175.

## 2.2. Proposed Parameter Combination-Driven Estimation Method

### 2.2.1. Basic Concepts and Assumptions

PCDE is a novel method for microparameter estimation. This method has two distinctive characteristics compared to LSE: 1) the allowance of one-on-one mapping between early (e.g., $\leq$ 10 min PI) and late (e.g., > 10 min PI) dynamics in TACs by limiting the precision of the estimated kinetic parameter (e.g., up to 2$^{nd}$ decimal place), and 2) employment of multi-aspect time-activity curve (TAC) in selection of best fits. We elaborate more on these next.

The first characteristic is based on two assumptions: 1) each microparameter has a finite range[8,16], and 2) the imaging system has a finite level of precision in determination of a microparameter (i.e., step size of a microparameter). Under these assumptions, only a finite number of TACs are available for a given range and precision, which enables to improve the probability of having one-on-one relationship between early and late dynamics by filtering out similar TACs. Indeed, with a kinetic parameter precision of 2$^{nd}$ decimal place (i.e., step size: 0.01), almost all TACs from 2TCM are likely to be unique and thus have a higher probability of correct one-to-one mapping between early and late dynamics for TACs. This improved uniqueness enables to predict a full TAC (i.e., early + late) in situations where the early dynamics are missing.

The second characteristic is a finer and more consistent comparison between the measured and reference truth TACs, compared to LSE. Inherently, the sum of squared error (SSE) cannot account for positive and negative errors differently[20–24]. Therefore, minimizing the SSE of concentration/activity (i.e., LSE) might not capture very small TAC trends well; something critical for microparameter estimation. Instead, other aspects of TAC (e.g., its 1$^{st}$ and 2$^{nd}$ derivatives) can be effective criteria for further finely assessing curve trends. Additionally, a comprehensive comparison of various aspects of TACs would yield more stable and balanced results. Relying solely on a single aspect for comparison could lead to significantly varied and unstable outcomes, influenced by factors such as noise level, type, number of passes in WB scans, measurement time intervals, and voxel positions within the body[25–27]. Thus, a comprehensive consideration of the multiple aspects of TAC would allow for a more consistent comparison. The details are presented in the next section.

2.2.2. Workflow and Similarity Measure

The workflow of the proposed method comprises three steps: 1) building a reference truth TAC database by setting microparameter range and precision of estimated parameters, 2) selecting the top-300 optimal parameter combinations with respect to SSE in ascending order, and subsequently, selecting the top-10 using the absolute difference of area under the curve (AUC) between the measured and ground truths in ascending order, and finally 3) selecting the best parameter combination using a comprehensive comparison based on multiple TAC aspects. Figure 1 shows the workflow of the proposed method.

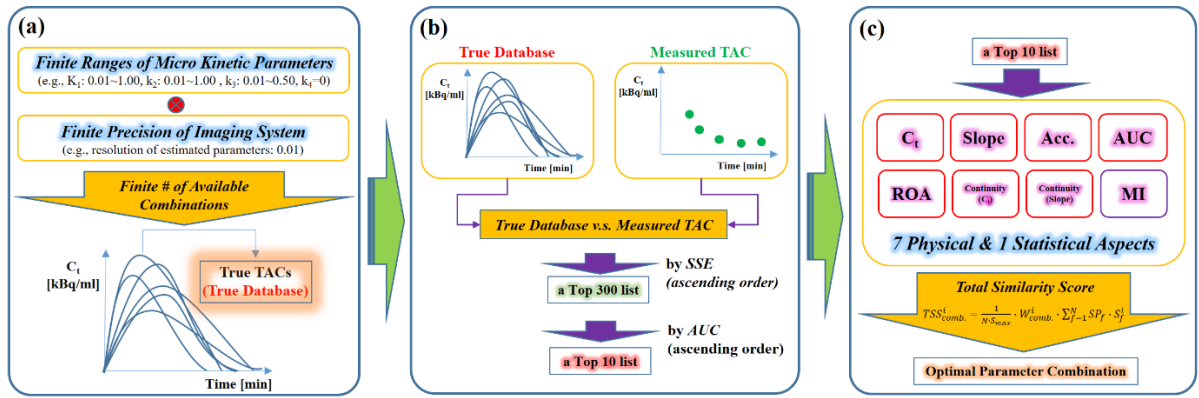

**Figure 1. PCDE workflow. (a): Building a reference truth TAC database. (b): Selecting the top 300 combinations followed by top 10 by comparing measured and reference TAC databases. (c): Selecting the optimal combination using comprehensive comparison based on multi-aspect of TAC.**

For the comprehensive comparison, towards picking best parameter combination as the one arriving at TAC with highest total similarity score (TSS), TSS was defined as follows:

$$TSS_{comb.}^i = \frac{1}{N \cdot S_{max}} \cdot W_{comb.}^i \cdot \sum_{f=1}^{N} SP_f \cdot S_f^i \qquad (3)$$

where i, f, and N denote an index for a parameter combination in the top-10 list, an index for an aspect of TAC, and the total number of aspects considered, respectively. $W_{comb.}^i$, $SP_f$, $S_f^i$, and

$S_{max}$ represent the relative weight of the i[th] combination, selection power for aspect f, a scaled score of the i[th] combination for aspect f, and the maximum scaled score, respectively (further elaborated in subsequent paragraphs). Table 5 shows the eight similarity metrics (aspects considered) and the order (ascending vs. descending) for assigning the scaled scores to each parameter combination set. Depending on the raw score ranking in the top-10 list, scaled scores for each combination were assigned from 10 to 1 in descending order (i.e., maximum score: 10, step size: 1).

**Table 5. Similarity measure and order to assign scaled scores to each combination.**

| Aspect | Similarity Metric | Order |
|---|---|---|
| $C_t$ | SSE | Ascending |
| Slope | SSE | Ascending |
| Acc. | SSE | Ascending |
| AUC | AD | Ascending |
| ROA | Itself | Descending |
| Continuity ($C_t$) | SE | Ascending |
| Continuity (slope) | SE | Ascending |
| MI | Itself | Descending |

$C_t$, concentration; Acc., acceleration; ROA, ratio of overlapped area; MI, mutual information; SSE, sum of squared error; SE, squared error; AD, absolute difference; Scaled score: 10 to 1 depending on the ranking among the top-10 lists (step size: 1).

As seen in Table 5, in our work we considered 8 physical and statistical aspects of TAC: 1) concentration/activity, 2) slope and 3) acceleration of TAC to consider a fine TAC trend, 4) AUC, 5) ratio of overlapped area (ROA) to compensate for a limitation of simple AUC comparison, continuities of 6) concentration and 7) slope at the earliest measurement time between true and measured quantities for each to account for the relatively higher importance of data at an early time after injection, and 8) mutual information as a statistical similarity measure[28,29].

In addition, to quantitatively account for the different capabilities of each TAC aspect in how well an aspect can distinguish parameter combinations in the top-10 list separately, we defined the relative selection power ($SP_f$) as follows:

$$SP_f \equiv \frac{CV_f}{\sum_{f=1}^{N} CV_f} \quad (4)$$

where f and N denote the index for an aspect of the TAC and total number of aspects considered, respectively, and $CV_f$ represents the coefficient of variation for aspect f. Supplemental Figure 1 shows the calculation process for the relative selection powers.

Furthermore, we defined a parameter combination weight (i.e., $W^i_{comb.}$) to account for the relative occurrence probability of a parameter combination in the top-10 list so that the more probable combination can contribute more to the TSS, assuming that each microparameter is independent of the others. The formulas are as follows:

$$W^i_{comb.} \equiv \frac{P^i_{comb.}}{\sum_{i=1}^{10} P^{i.}_{comb.}} \quad (5)$$

$$P^i_{comb.} \equiv P^{K1 \, of \, ith \, comb.}_{K1} \cdot P^{k_2 \, of \, ith \, comb.}_{k2} \cdot P^{k_3 \, of \, ith \, comb.}_{k3} \quad (6)$$

where i denotes an index for a parameter combination, and $P^{K1 \, of \, ith \, comb.}_{K1}$, $P^{k_2 \, of \, ith \, comb.}_{k2}$ and $P^{k_3 \, of \, ith \, comb.}_{k3}$ represent the probabilities of having $K_1$, $k_2$, $k_3$ for the i$^{th}$ combination, respectively. $P^i_{comb.}$ is the probability of occurrence of the i$^{th}$ combination. Supplemental Figure 2 shows the calculation process for the parameter combination weights.

2.3. Kinetic Parameters of Interest for Comparison Study

On the noisy virtual dynamic dataset, kinetic modeling was performed through each method (i.e., LSE and PCDE), and the kinetic parameters of interest for comparison are defined as follows.

2.3.1. Kinetic Microparameters.

For the microparameters, we compared the LSE-based 2TCM[30], implemented via the Levenberg-Marquardt (LM) algorithm (function tolerance: 10$^{-9}$, max iterations: 1000), against the proposed PCDE method. Because we focused on the irreversible uptake process, only parametric $K_1$, $k_2$, and $k_3$ images were compared.

2.3.2. Kinetic Macroparameters

For the macroparameters, we compared the parametric images of the LSE-based Patlak graphical analysis (PGA)[30,31] with those of PCDE. Assuming an irreversible or nearly irreversible uptake process in 2TCM (i.e., $k_4 \approx 0$), the PGA formula can be derived as follows:

$$\frac{C_{PET}(t)}{C_p(t)} = K_i \cdot \frac{\int_0^t C_p(\tau)d\tau}{C_p(t)} + V_d, \quad t > t^* \quad (7)$$

where $t^*$ denotes the time required to reach equilibrium between the plasma and the first compartment in the 2TCM.

Furthermore, assuming that the blood volume fraction was negligible (i.e., $V_b \approx 0$), we defined the net influx rate constant $K_i$ and volume of distribution $V_d$ as follows:

$$K_i = \frac{K_1 k_3}{k_2 + k_3} \quad (8)$$

$$V_d = \frac{K_1 k_2}{(k_2 + k_3)^2} \tag{9}$$

2.4. Quantitative Evaluation Criteria

2.4.1. General Image Quality

*Normalized Bias (NBias)*. As a measure of accuracy, NBias is determined by first calculating NBias$_i$ for the i$^{th}$ voxel of an ROI over all R noise realizations and subsequently averaging over all voxels of that ROI as follows:

$$\text{NBias} = \frac{1}{n}\sum_{i=1}^{n}\left(\frac{|\overline{f_i} - \mu_i|}{\mu_i}\right) = \frac{1}{n}\sum_{i=1}^{n} NBias_i \tag{10}$$

where $\overline{f_i} = (1/R)\sum_{r=1}^{R} f_i^r$; $f_i^r$ denotes the i$^{th}$ voxel value from the r$^{th}$ noise realization, and $\mu_i$, n, and R represent the truth of the i$^{th}$ voxel, the number of voxels in an ROI, and the number of noise realizations, respectively.

*Normalized Standard Deviation (NSD)*. As a precision measure, the NSD$_i$ of the i$^{th}$ voxel was first calculated over all R realizations, followed by averaging over all n voxels of an ROI to calculate the NSD of the ROI as follows:

$$\text{NSD} = \frac{1}{n}\sum_{i=1}^{n}\frac{\sqrt{\frac{1}{R-1}\sum_{r=1}^{R}(f_i^r - \overline{f_i})^2}}{\overline{f_i}} = \frac{1}{n}\sum_{i=1}^{n} NSD_i \tag{11}$$

*Normalized Root Mean Squared Error (NRMSE)*. As a measure of comprehensive performance (i.e., combined measure of accuracy and precision), NRMSE$_i$ was first calculated for each i$^{th}$ voxel over all realizations, followed by spatial averaging over all voxels of an ROI to calculate the NMSE for an ROI as follows:

$$\text{NRMSE} = \frac{1}{n}\sum_{i=1}^{n}\frac{\sqrt{\frac{1}{R}\sum_{r=1}^{R}(f_i^r - \mu_i)^2}}{\mu_i} = \frac{1}{n}\sum_{i=1}^{n} NRMSE_i \tag{12}$$

For each ROI of interest (Table 1), the calculations of all three quantities were repeated by changing the number of OSEM iterations, as listed in Table 4. To compare the general image quality between each estimation method (i.e., LSE vs. PCDE), we plotted the NBias-NSD trade-off curves. In addition, NRMSEs were plotted against the number of iterations.

2.4.2. Overall Visibility and Tumor Detectability

*Signal-to-Noise Ratio (SNR)*. As a measure of the overall visibility relevant to the identification of suspicious lesions in WB (i.e., global inspection), the SNR of an ROI was determined by averaging the SNRs over all noise realizations as follows:

$$\text{SNR} = \frac{1}{R}\sum_{r=1}^{R} \frac{\bar{f}_r}{\sqrt{\frac{1}{n-1}\sum_{i=1}^{n}(f_i^r - \bar{f}_r)^2}} = \frac{1}{R}\sum_{r=1}^{R} SNR_r \quad (13)$$

where $\bar{f}_r = (1/n)\sum_{i=1}^{n} f_i^r$.

*Spatial Noise (NSD$_{spatial}$).* As another measure of overall visibility, the NSD$_{spatial}$ of an ROI was calculated by averaging the NSDs over all realizations as follows:

$$NSD_{spatial} = \frac{1}{R}\sum_{r=1}^{R} \frac{\sqrt{\frac{1}{n-1}\sum_{i=1}^{n}(f_i^r - \bar{f}_r)^2}}{\bar{f}_r} = \frac{1}{R}\sum_{r=1}^{R} NSD_{spatial}^r \quad (14)$$

By comparing Equations (11) and (14), it should be noted that NSD quantifies the average level of noise across multiple realizations at each voxel for an ROI, whereas NSD$_{spatial}$ known as ROI roughness, measures the average of the spatial noise across multiple realizations for an ROI[8].

*Tumor-to-Background Ratio (TBR).* As a measure of tumor detectability within a particular organ (i.e., local inspection), TBR was determined as follows:

$$\text{TBR} = \frac{1}{R}\sum_{r=1}^{R} \frac{\overline{f_r^{Tumor}}}{\overline{f_r^{BKG.}}} = \frac{1}{R}\sum_{r=1}^{R} TBR_r \quad (15)$$

where $\overline{f_r^{Tumor}}$ and $\overline{f_r^{BKG.}}$ denote $\bar{f}_r$ of tumor and background ROI, respectively.

*Contrast-to-Noise Ratio (CNR).* As a measure of tumor detectability within a specific organ (i.e., local inspection), the CNR was calculated as follows:

$$\text{CNR} = \frac{1}{R}\sum_{r=1}^{R} \frac{|\overline{f_r^{Tumor}} - \overline{f_r^{BKG.}}|}{\sigma_r^{BKG.}} = \frac{1}{R}\sum_{r=1}^{R} CNR_r \quad (16)$$

where $\sigma_r^{BKG.} = \sqrt{\frac{1}{n-1}\sum_{i=1}^{n}(f_i^r - \overline{f_r^{BKG.}})^2}$

*Relative Error of TBR (RE$_{TBR}$).* It is possible to have a misleading (i.e., erroneously higher) TBR and/or CNR originating from a high bias (i.e., the wrongly increased/decreased mean ROI) and/or zero-like noise (i.e., the noise is approximately zero), owing to the local minimum issue of the LSE. Hence, the RE$_{TBR}$ was also calculated as an auxiliary measure.

$$RE_{TBR} = \frac{|TBR_{Measured} - TBR_{Truth}|}{TBR_{Truth}} \quad (17)$$

where $TBR_{Measured}$ and $TBR_{Truth}$ denote a measured and true TBR, respectively.

2.4.3. Overall Performance Metrics

To verify the overall performance of each parametric image, the overall NBias, NSD, NRMSE, SNR, and NSD$_{spatial}$ metrics were defined as the volume-weighted averages of the individual ROIs metrics[8].

2.5. Patient Study

In addition to performance testing on virtual dynamic datasets, the proposed method was also implemented on actual patient datasets to further verify clinical applicability. In this work, we focus on anecdotal study for initial assessment, and a large-scale patient study is topic of an upcoming study (see discussion). Table 6 summarizes scanning protocols for two patients undergoing $^{18}$F-DCFPyL PET scans, involving prostate-specific membrane antigen (PSMA) targeted imaging.

The quantitative evaluation of parametric images includes the analysis of: 1) overall visibility relevant to the identification of suspicious lesions in WB (i.e., SNR$_{overall}$), and 2) overall lesion detectability (i.e., CNR$_{overall}$, TBR$_{overall}$). The ROIs of all lesions for each patient were defined and confirmed by a nuclear medicine physician.

**Table 6. Dynamic scanning protocols for $^{18}$F-DCFPyL.**

| Item | Patient #1 | Patient #2 |
|---|---|---|
| Injected activity | 9.14 mCi | 7.16 mCi |
| Scanner | GE Discovery MI | GE Discovery MI |
| Dimensions | 256 x 256 x 409 | 256 x 256 x 409 |
| Voxel size | 2.73 x 2.73 x 2.8 mm$^3$ | 2.73 x 2.73 x 2.8 mm$^3$ |
| Total acquisition time (cardiac + WB) | 87 min | 92 min |
| Image acquisition for WB | 7-87 min | 9-92 min |
| Time interval | 5 min | 5 min |
| # of passes | 16 | 16 |
| # of beds | 6 | 6 |

3. Results

3.1. NBias-NSD Trade-off Curves and NRMSE Results

3.1.1. Kinetic Microparameters

Supplemental Figures 3, 4, and 5 show the ROI-based NBias-NSD trade-off curves and NRMSE results for the parametric $K_1$, $k_2$, and $k_3$ images, respectively. Overall the proposed PCDE method showed lower NBias and NSD compared to the LSE-based 2TCM, which allows much lower

NRMSEs for all normal WB organs of interest; the common standard shows smaller NSDs in $K_1$ images. However, significantly high levels of NBias result in larger NRMSEs for all ROIs.

Figure 2 shows the overall NBias-NSD trade-off curves and NRMSE results. At five OSEM iterations, using our PCDE method, the overall NRMSEs were considerably reduced by 57.5, 71.1, and 56.1 [%] in the parametric $K_1$, $k_2$, and $k_3$ images, respectively.

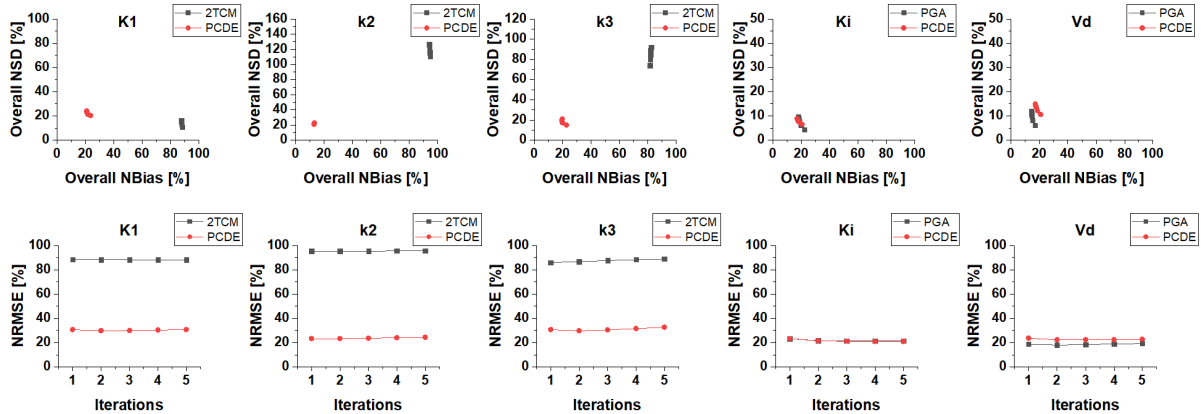

Figure 2. Overall NBias-NSD trade-off curves (i.e., first row) and NRMSE results with increasing OSEM iterations (i.e., second row) for each parametric image. Microparameters: first three columns; Macroparameters: last two columns.

### 3.1.2. Kinetic Macroparameters

Supplemental Figures 6 and 7 show the ROI-based NBias-NSD trade-off curves and NRMSE results for the parametric $K_i$ and $V_d$ images, respectively. No significant differences between the LSE-based PGA and PCDE were observed. For $V_d$, the PGA shows a slightly better performance, but the differences are less than 10 [%] in most cases.

Figure 2 shows the overall NBias-NSD trade-off curves and NRMSE results. At five OSEM iterations, using our proposed PCDE method, the overall NRMSE for $K_i$ was reduced by 0.4 [%]. However, the overall NRMSE for $V_d$ was increased by 3.3 [%], indicating no significant difference between the two methods.

## 3.2. Overall Visibility and Tumor Detectability

### 3.2.1. Kinetic Microparameters

The first three columns of Figure 3 shows the overall visibility results for the parametric $K_1$, $k_2$, and $k_3$ images. After five OSEM iterations, the overall SNR increased by 0.2, 4.1, and 2.4, and the overall $NSD_{spatial}$ decreased by 0.2, 5.4, and 4.1 for the parametric $K_1$, $k_2$, and $k_3$ images, respectively, indicating excellent performance of our proposed method in both aspects simultaneously.

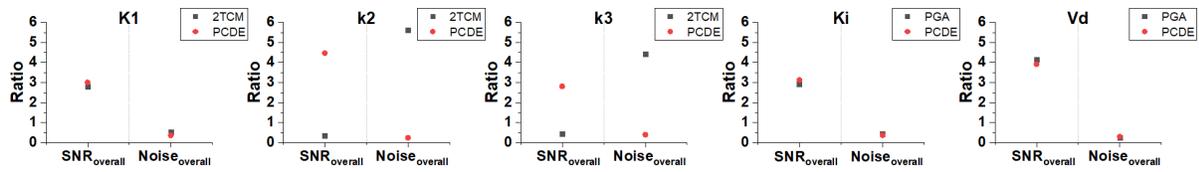

**Figure 3.** Overall visibility in each parametric image. Microparameters: first three columns; Macroparameters: last two columns. (OSEM iterations=5). Matrices: overall SNR and $NSD_{spatial}$.

The first three columns of Figure 4 show the tumor detectability results for each tumor in the parametric $K_1$, $k_2$, and $k_3$ images. After five OSEM iterations, although there was no clear improvement in CNR in the $k_2$ images from the proposed method, the CNR for a lung tumor increased by 1.3 and 1.0, and that for a liver tumor increased by 1.2, and 9.8 in the $K_1$ and $k_3$ images, respectively. In addition, the $RE_{TBR}$ of a lung tumor decreased by 17.5, 82.2, and 68.4, and that of the liver tumor decreased by 255.8, 1733.5, and 80.3 [%] in the $K_1$, $k_2$, and $k_3$ images, respectively. Figure 5 shows an example of microparametric images.

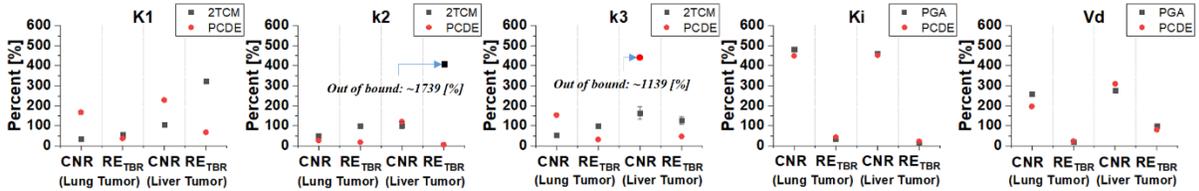

**Figure 4.** Tumor detectability in each parametric image. Microparameters: first three columns; Macroparameters: last two columns. Matrices: CNR [%] and $RE_{TBR}$ [%]. (OSEM iterations=5).

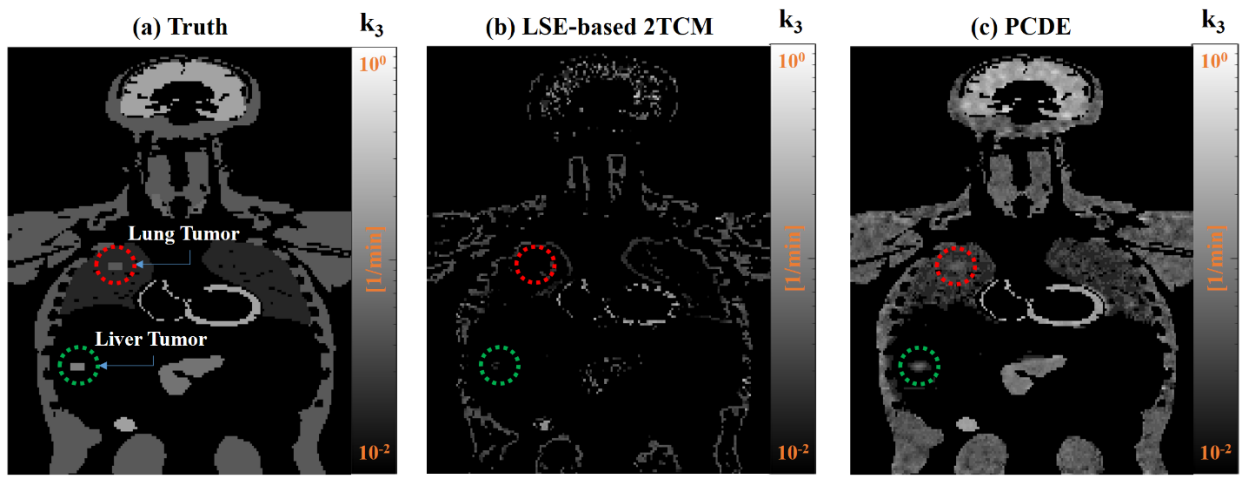

**Figure 5.** Parametric $k_3$ images with five OSEM iterations. (a): Ground Truth. (b): LSE-based 2TCM. (c): PCDE. (OSEM iterations=5, Noise Realization index=1).

3.2.2. Kinetic Macroparameters

The last two columns of the Figure 3 show the overall visibility results for the parametric $K_i$ and $V_d$ images. There were no substantial differences between the two methods in either aspect. The last two columns of Figure 4 show the tumor detectability results for each tumor in the parametric $K_i$ and $V_d$ images. For both tumors, the differences in CNR were within 0.5, and the differences in $RE_{TBR}$ were within 10 [%] in most cases, except for the case with a decrease in $RE_{TBR}$ by 19.6 [%] for a liver tumor in the $V_d$ images using the proposed method. Figure 6 shows an example of macroparametric images.

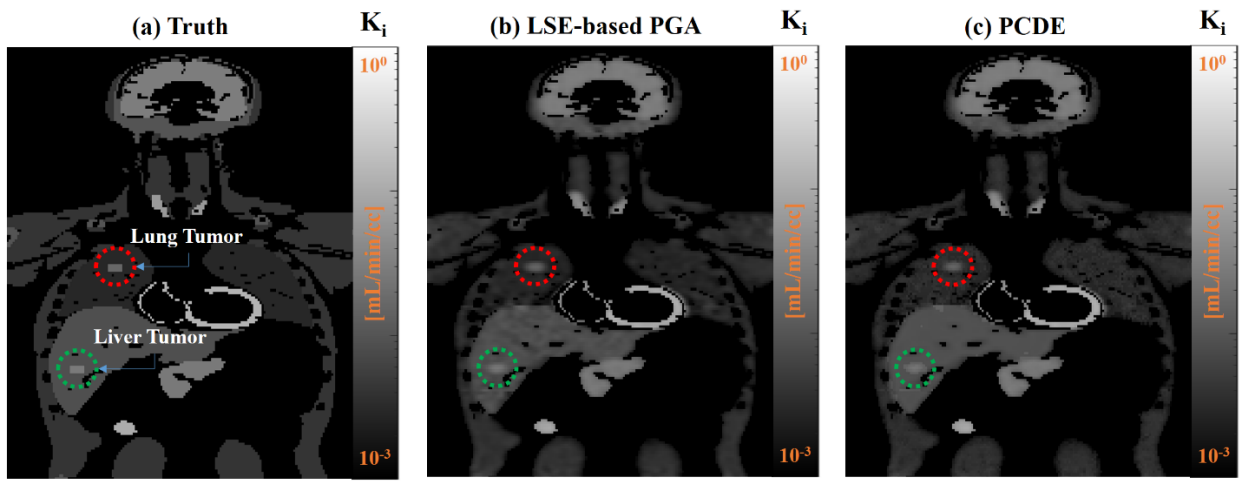

**Figure 6.** Parametric $K_i$ images with five OSEM iterations. (a): Ground Truth. (b): LSE-based PGA. (c): PCDE. (OSEM iterations=5, Noise Realization index=1).

3.3. Patient Study

### 3.3.1. Overall Visibility

The first three columns of Figure 7 show the overall visibility results of $^{18}$F-DCFPyL for microparametric images. The averaged overall SNR (i.e., average of individual patient's metric) increased by 1.19±0.25, 2.06±0.42, and 0.80±0.16 for the parametric $K_1$, $k_2$, and $k_3$ images, respectively. Each first row of the Figures 8 and 9 shows examples of micro-parametric images for each patient. Overall, compared to 2TCM, better definitions with less noise via PCDE in all micro-parametric images were verified.

In addition, the last two columns of Figure 7 show the overall visibility for macroparametric images. The averaged overall SNR increased by 0.09±0.03 and 0.37±0.58 for the parametric $K_i$ and $V_d$ images, respectively. Each second row of the Figures 8 and 9 shows examples of macro-parametric images for each patient. Overall, there were no visually significant differences between two methods (i.e., PGA vs. PCDE).

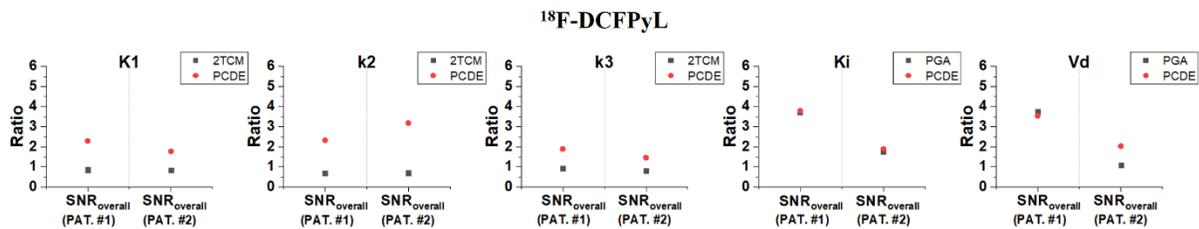

Figure 7. Overall visibility in each parametric image. Micro-parameters: first three columns; Macro-parameters: last two columns. PAT.=patient. Radiotracer: $^{18}$F-DCFPyL.

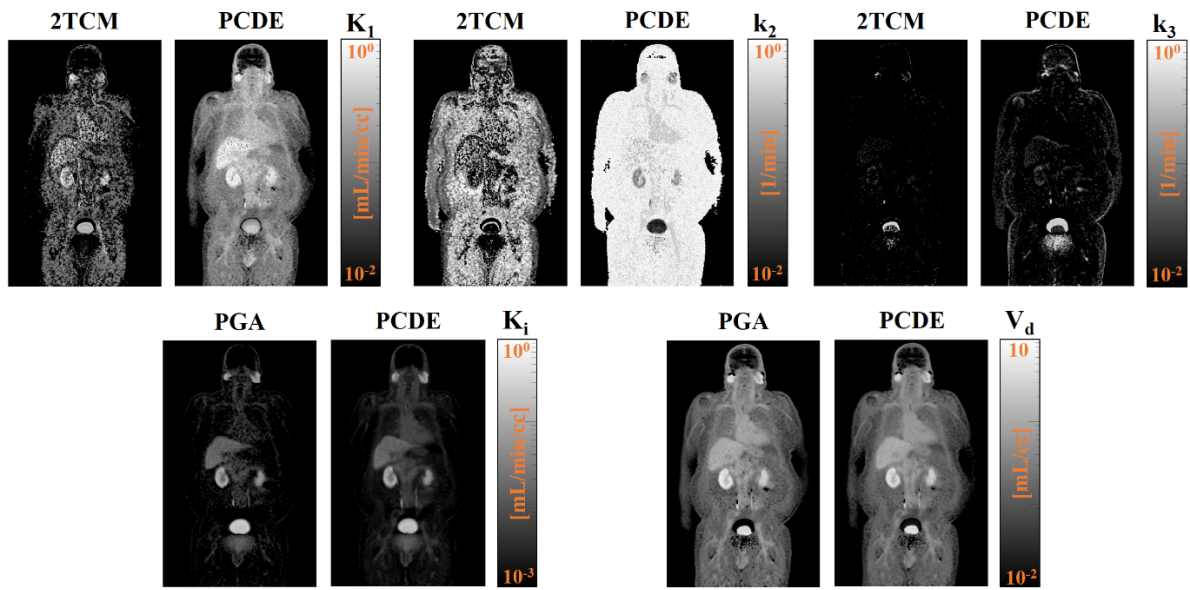

Figure 8. Example of parametric images focusing on overall visibility. Micro-parameters ($K_1$, $k_2$, and $k_3$): the first row. Macro-parameters ($K_i$ and $V_d$): the second row. Radiotracer: $^{18}$F-DCFPyL (Patient #1). We showed results for conventional 2TCM approach vs. our proposed PCDE approach.

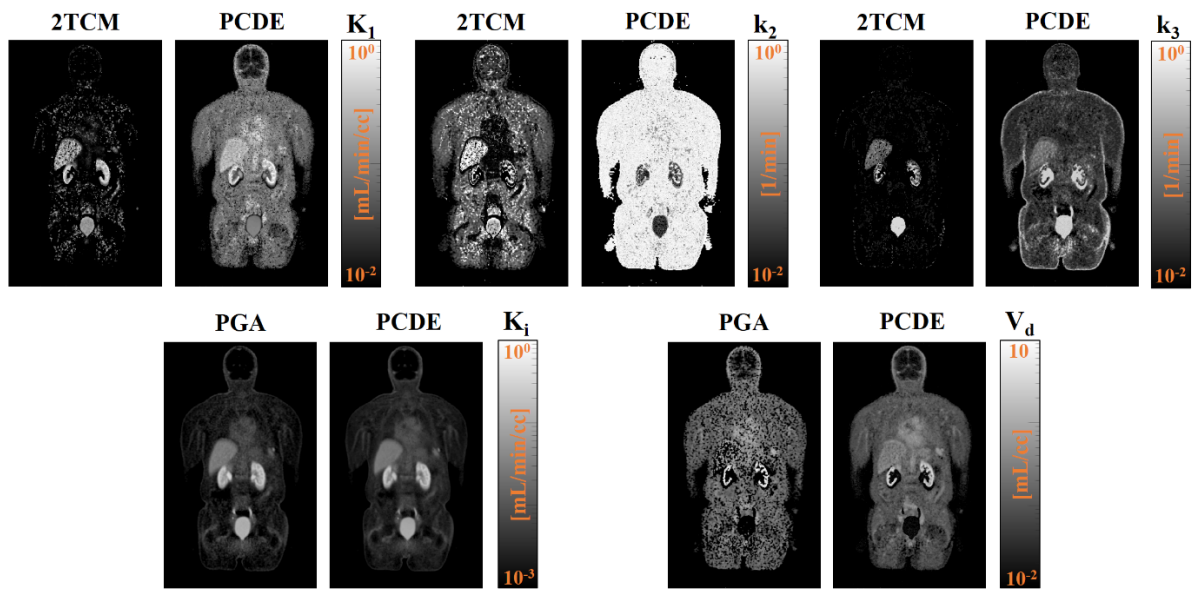

**Figure 9.** Example of parametric images focusing on overall visibility. Micro-parameters ($K_1$, $k_2$, and $k_3$): the first row. Macro-parameters ($K_i$ and $V_d$): the second row. Radiotracer: $^{18}$F-DCFPyL (Patient #2). We showed results for conventional 2TCM approach vs. our proposed PCDE approach.

3.3.2. Lesion Detectability

The first three columns of Figure 10 show the tumor detectability for microparametric images. The overall CNR increased by 2.54, 1.99, and 1.29, and the overall TBR increased by 1.21, 0.39, and 1.84 for the parametric $K_1$, $k_2$, and $k_3$ images, respectively, indicating excellent performance of our proposed method in both aspects simultaneously. For instance, Figure 11 presents examples of parametric $K_1$ and $k_3$ images. Compared to reference method (i.e., 2TCM), the enhanced lesion detectability was verified via PCDE.

In addition, the last two columns of Figure 10 show the tumor detectability for macroparametric images. The overall CNR increased by 0.21 and -0.06, and the overall TBR increased by -0.49 and 0.16 for the parametric $K_i$ and $V_d$ images, respectively. The minus sign represents the decrease in metric of interest.

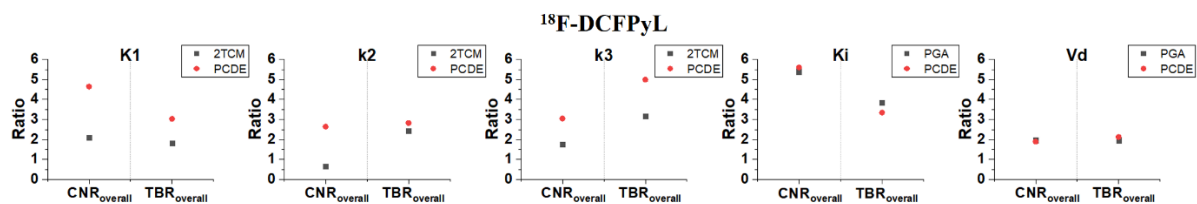

**Figure 10.** Lesion detectability in each parametric image. Micro-parameters: first three columns; Macro-parameters: last two columns. Matrices: $CNR_{overall}$ and $TBR_{overall}$ (the # of lesions: 2). Radiotracer: $^{18}$F-DCFPyL.

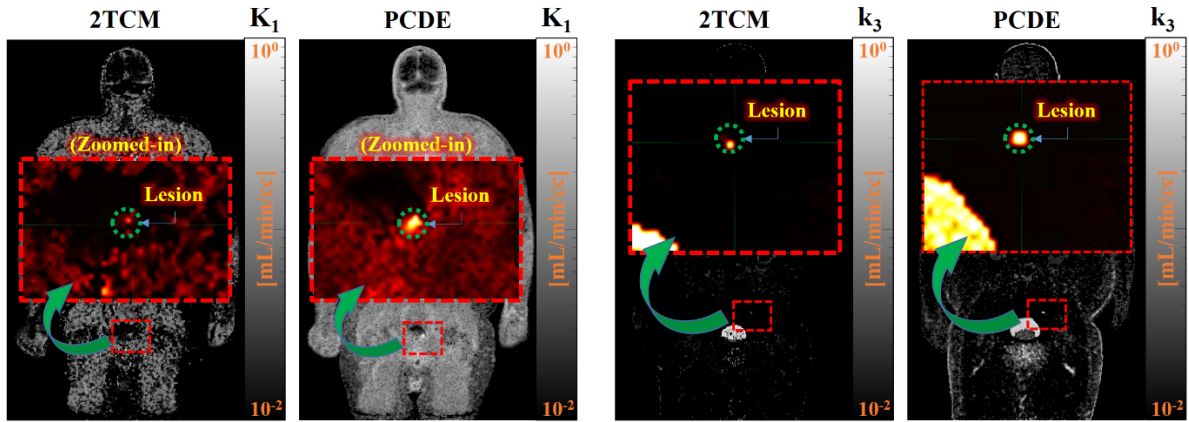

Figure 11. Examples of parametric images focusing on lesion detectability. Micro-parameter: $K_1$ (lesion #1) and $k_3$ (lesion #2). Radiotracer: $^{18}$F-DCFPyL (Patient #1).

## 4. Discussion

This study introduces our proposed method (i.e., PCDE) and compares it to the common standard parameter estimation method for kinetic modeling invoking LSE. The comparison study was performed on virtual dynamic dataset and focusing on two aspects:1) general image quality for major normal organs in WB, and 2) overall visibility and tumor detectability.

First, we verified that PCDE could improve the quality of microparametric images (i.e., NBias, NSD, and NRMSE). For the $K_1$ image, the LSE-based 2TCM showed better results in terms of NSD. However, the considerably higher level of bias compared to PCDE resulted in a larger NRMSE, reducing the overall performance compared to PCDE. Moreover, because multiple local minima can cause variability (e.g., NSD or $NSD_{spatial}$) with high bias, the lower level of NSD from the LSE-based 2TCM could be due to the local minimum issue of LSE[32–34] instead of the actual benefit of LSE for $K_1$ images. Supplemental Figure 8 shows the example of an erroneously lower level of $NSD_{spatial}$ with high bias in the $K_1$ image generated from the LSE-based 2TCM. When considering that NBias through the LSE-based 2TCM show an extremely high bias (i.e., 96.6 [%]), we can indirectly expect that the lower level of $NSD_{spatial}$ is caused by the local minimum issue rather than the improved performance of LSE.

For macroparameters, there was no significant difference between the PCDE and LSE-based PGA. This was expected because the relative benefit of PCDE compared to the reference (i.e., LSE-based PGA) would not be significant because the macroparameter estimation from the reference method already has good accuracy (i.e., NBias) and precision (i.e., NSD) owing to the linearized fit-type function for PGA[30,31].

In addition, we verified the improved overall visibility (i.e., overall SNR, overall $NSD_{spatial}$) and tumor detectability (i.e., CNR, $RE_{TBR}$) in the microparametric images, except for CNRs in $k_2$ images. For $k_2$ images, there was a negligible difference between the two methods (i.e., ≤ 0.5). However, a high positive bias of the tumor, high negative bias of the background, and erroneously zero-like $NSD_{spatial}$ originating from the local minimum issue of LSE may highly mislead CNR value (i.e., erroneously high CNR), which cannot provide any actual benefit for tumor detectability on images. Thus, a comparison based solely on CNR may lead to incorrect conclusions regarding tumor detection capability.

Supplemental Figure 9 shows the example of a misleading CNR and the necessity of $RE_{TBR}$ for a fair comparison in this simulation study. Even though the CNR from the reference shows a slightly better CNR than that of PCDE (i.e., $CNR_{ref}$=1.8, $CNR_{PCDE}$=1.5), there is no actual relative benefit from the reference method in terms of tumor detection. Moreover, the relatively better CNR originates from high levels of bias in the liver tumor and background (i.e., highly negative bias) as shown in the figure.

Therefore, in this study, we included $RE_{TBR}$ as an auxiliary measure to minimize the possibility of incorrect conclusions regarding tumor detection capability. Considering that PCDE showed much lower $RE_{TBR}$ values even for cases where the CNRs were quite similar (due to the misleading CNR from the reference method), we expect improved tumor detectability through PCDE compared to that of the reference.

With microparametric images from PCDE (e.g., Figure 5), improving the overall SNR and $NSD_{spatial}$ would help identify suspicious regions in WB globally (i.e., global inspection). The improved CNR and $RE_{TBR}$ performance would directly lead to improved tumor detectability locally within a particular organ (i.e., local inspection). For macroparameters (e.g., Figure 6), there were no significant differences in the overall visibility and tumor detectability between the two methods. This is understandable because the two methods had no significant differences in general image quality (i.e., NBias, NSD, and NRMSE).

Overall, our proposed PCDE method provides enhanced microparametric images on not only virtual dynamic datasets but also real patient data, supporting application to clinical studies, and need for more exhaustive studies.

4.1 Comparison with other studies

Our study contributes to enable reliable WB kinetic modeling in regular-axial field-of-view PET scanners (i.e., multi-pass protocols on a limited axial FOV), tackling 3 important points (as elaborated next): 1) minimization of adverse effects in previously proposed techniques, 2) potential applicability for shorter scan durations, and 3) avoidance of the local minimum issues discussed above.

For the first point, the protocol proposed by Karakatsanis et al.[8,14] was optimized based on macroparametric images (i.e., $K_i$) and was used 6 min after injection to scan the cardiac region. Because the macroparameters of PGA only require data after the mechanism reaches kinetic equilibrium[8,35,36], the loss of early dynamics of TAC would not adversely affect parameter estimation. However, unlike macroparameters, early dynamics are critical for microparameter estimation because they typically include near-peak data considerably influenced by microparameter combinations. Although the accuracy and precision of the microparameter estimation need to be improved further relative to those of the macroparameter (i.e., Figure 2), it offers increased improvements for each microparameter compared to the common standard. This indicates a substantial reduction in the adverse effects of the protocol favorably optimized for macroparameter estimation.

In addition, for the second point, the comprehensive comparison based on the multi-aspect of TAC can offer more stabilized parameter estimation (i.e., less variation of performance) from various image acquisition-related factors (e.g., the number of passes, time interval, voxel position, noise level, and type), compared to the case considering only one single factor (e.g., SSE for LSE). Therefore, we expect our proposed method to perform better even when using a dynamic PET dataset scanned only for 30 min, realistically achieving the shortest scan duration for a typical PET-based WB kinetic modeling for microparameter estimation.

All results reported in this study are based on a simulated dynamic dataset scanned only 40 min PI, which is 5 min shorter than the optimal acquisition length suggested by Dr. Karakatsanis (i.e., 45 min) and 20 min shorter than the typical time required for dynamic PET acquisition[1] for kinetic modeling (i.e., 60 min). Hence, we can expect the promising applicability of the proposed method to studies involving shorter scan duration.

Moreover, the PCDE avoids the local minimum issue by systematically evaluating various aspects of TAC and selecting the best parameter combination, rather than relying on an iterative approach to find an optimal value. Consequently, unlike the LSE method, the PCDE does not necessitate an initial guess for parameter estimation. However, PCDE also uses curve fitting to model a measured TAC, but the later dynamics of TAC (e.g., >10 min after injection) can be well-fitted using a single exponential function (i.e., fit-type function: $c - a \cdot e^{-bt}$, fit parameter: a, b, c), which can be an automatic process without a manual initial guess because of its negligible dependence on the initial values.

Tackling the local minimum issue is critical for the active use of kinetic modeling in clinics for two reasons. First, it is not necessary to set starting points for each voxel (i.e., voxel-wise computation) or ROIs (i.e., ROI-based computation). Compared to curve fitting for the later dynamics of TAC (i.e., a single exponential shape), curve fitting for the entire dynamics of TAC (i.e., a surge-like shape) is most likely to have starting point dependency, especially if near-peak data are missing either partially or completely. Thus, for clinical use, starting points must be set subtlety through repetition to minimize the adverse effects of the local minimum issue (i.e., finding a global minimum), which is time-consuming when performed for each voxel or ROI, preventing the routine application of kinetic modeling in the clinic. Second, by minimizing the starting point dependency, the interpersonal error of the estimated kinetic parameters can be considerably reduced, which is critical for the consistency of kinetic modeling results and large-scale data comparison across different institutions worldwide.

4.2 Limitations

Some limitations in our proposed method indicate the need for further studies. First, the computational speed of PCDE is approximately $1.1 \times 10^{-3}$ s/voxel; therefore, approximately 2 hours of computing are needed to perform WB kinetic modeling for a typical active patient volume size in the clinic with typical hardware specifications (e.g., CPU: AMD Ryzen 9 5900HX, RAM: 32.0 GB, platform: MATLAB R2021b, precision of estimated parameter: 0.01). For use in routine practice, at least 100 times the current computational speed (i.e., ~$10^{-5}$ s/voxel) is needed to complete the computation in a few minutes. Parallelized computation using a graphical processing unit (GPU) will allow us to achieve this.

Secondly, in this study, we limited the maximum allowable value of the microparameter to 1 (for

$K_1$ and $k_2$) and 0.5 (for $k_3$), respectively. Although for $^{18}$F-FDG, almost all microparameters for each ROI in WB were within the desired ranges[8,16], we need to broaden the range to increase applicability to diverse types of radiotracers.

Thirdly, in the present work irreversible uptake process was assumed (i.e., $k_4 \approx 0$), as mentioned previously. Such an assumption is quite prevalent in past and ongoing studies with a number of radiopharmaceuticals (e.g. FDG, DCFPyL). Incorporating additional modeling of $k_4$ is certainly possible, and can result in more accurate estimates, but at the expense of reduced precision (increased noise) in parametric images given the extra degree of freedom in fitting. In any case, depending on a specific radiopharmaceutical of interest, such additional modeling can certainly be performed and studied.

Overall, with the addition of a reversible process and a broader range of parameters, we anticipate that the GPU-accelerated PCDE approach will enable the widespread use of typical PET-based WB kinetic modeling for kinetic microparameters. This method ensures both reasonable computational time and compatibility with various types of radiotracers.

Furthermore, despite significant improvements via PCDE, the overall levels of NBias and NSD tend to be beyond 10% (i.e., near 20%), and non-negligible variations among ROIs exist (e.g., supplemental Figures 1-7), implying that the proposed method may still be insufficient for use in routine practice. We expect that the exploitation of de-noising techniques such as the finite Legendre transform-based low-pass filter with excellent de-noising performance for the exponential type curve (i.e., typical shape of TAC after peak) without the phase shift[37] and/or noise propagation pattern learning through machine/deep learning algorithms (i.e., noise propagation from the sinogram domain into image domain) could reduce the overall levels of NBias and NSD within 10%. Moreover, it can reduce variations among ROIs (i.e., consideration of different noise propagation patterns at each position).

Finally, a validation study based on real patient data should be conducted. We are actively collecting patient data (e.g., Clinical Trial ID: NCT04017104) categorized by a specific tumor detection mechanism such as $^{18}$F-FDG by glucose metabolism[38], $^{18}$F-DCFPyL and $^{68}$Ga-HTK by targeting PSMA[39,40], and $^{18}$F-AmBF3 by targeting somatostatin receptor 2 (SSTR2)[41]. We expect to perform a validation study based on extensive patient data in the near future.

## 5. Conclusions

We compared the performance of kinetic parameter estimation between the common standard (LSE) and proposed PCDE method, focusing on general image quality, overall visibility, and tumor detectability. Although there were no significant differences in macroparameter estimation, significant improvements in the microparameters were demonstrated. PCDE can enable typical PET-based WB kinetic modeling for kinetic microparameters, which has been almost nonexistent owing to significant uncertainties in estimates when using LSE. Overall, our proposed framework enables microparametric imaging as applied to dynamic WB imaging protocols on regular-axial field-of-view PET scanners.

## Acknowledgments

We acknowledge the Natural Sciences and Engineering Research Council of Canada (NSERC) Discovery Grants RGPIN-2019-06467 and RGPIN-2021-02965, and the Canadian Institutes of Health Research (CIHR) Project Grant PJT-162216.We appreciate helpful discussions with Dr. Hamid Abdollahi. All figures and tables are courtesy of graduate thesis[42] at the University of British Columbia (UBC).

## Conflict of Interest Statement

No potential conflicts of interest relevant to this article were reported.

## Source Code Availability

The source code for PCDE is available for download from our GitHub software repository: https://github.com/leeronaldonew/PCDE

# Supplemental Figures

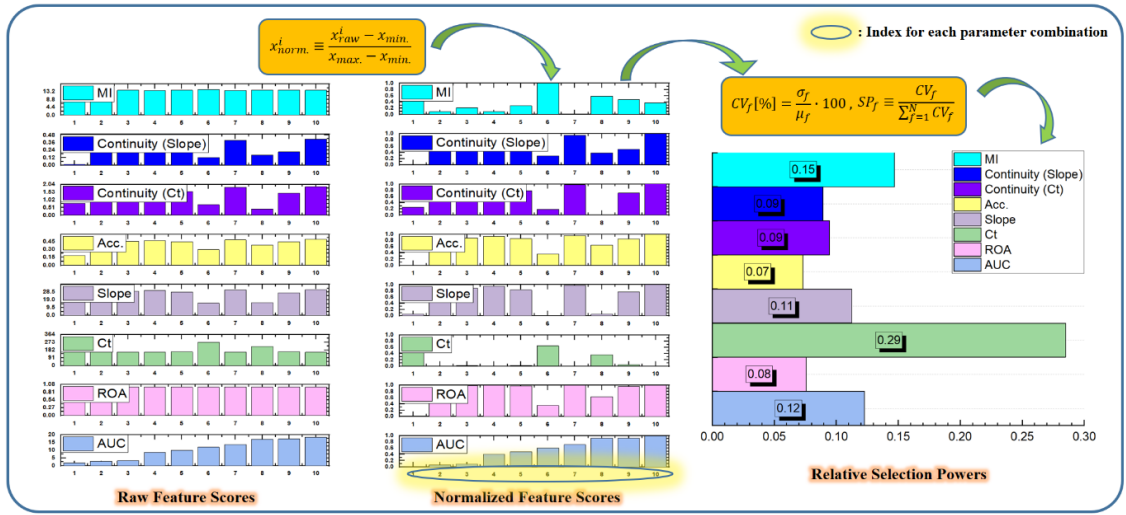

**Supplemental Figure 1.** Calculation of relative selection powers for each aspect of TAC. (1) Calculate normalized scores for each aspect by min-max normalization. (2) Calculate coefficients of variation for each aspect and the relative values that represent selection powers for each aspect.

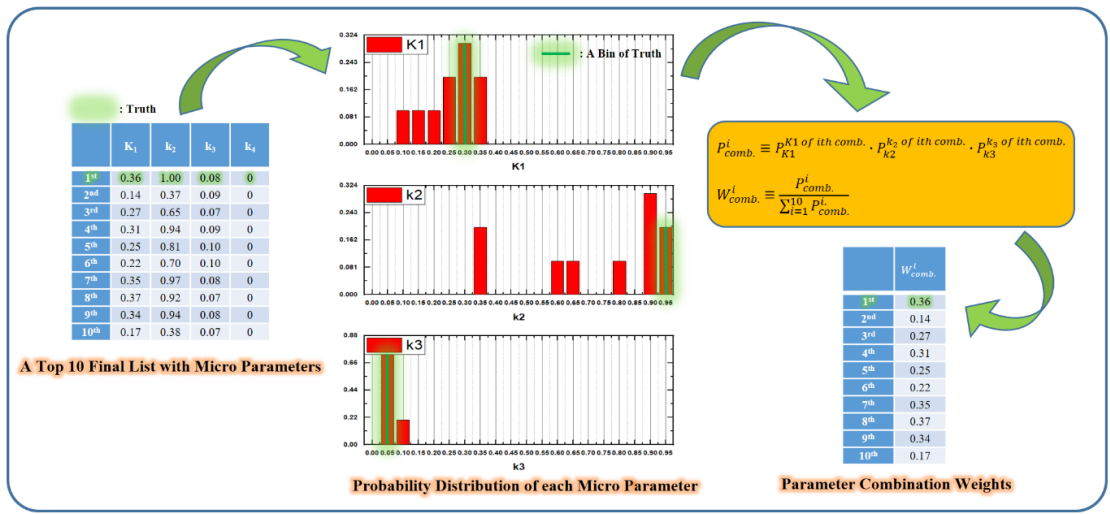

**Supplemental Figure 2.** Calculation of parameter combination weights for each combination in the top-10 list. (1) Acquire probability distributions of microparameters from the top 10 list. (2) Calculate occurrence probabilities for each combination in the top-10 list and relative values that represent the weights for each parameter combination.

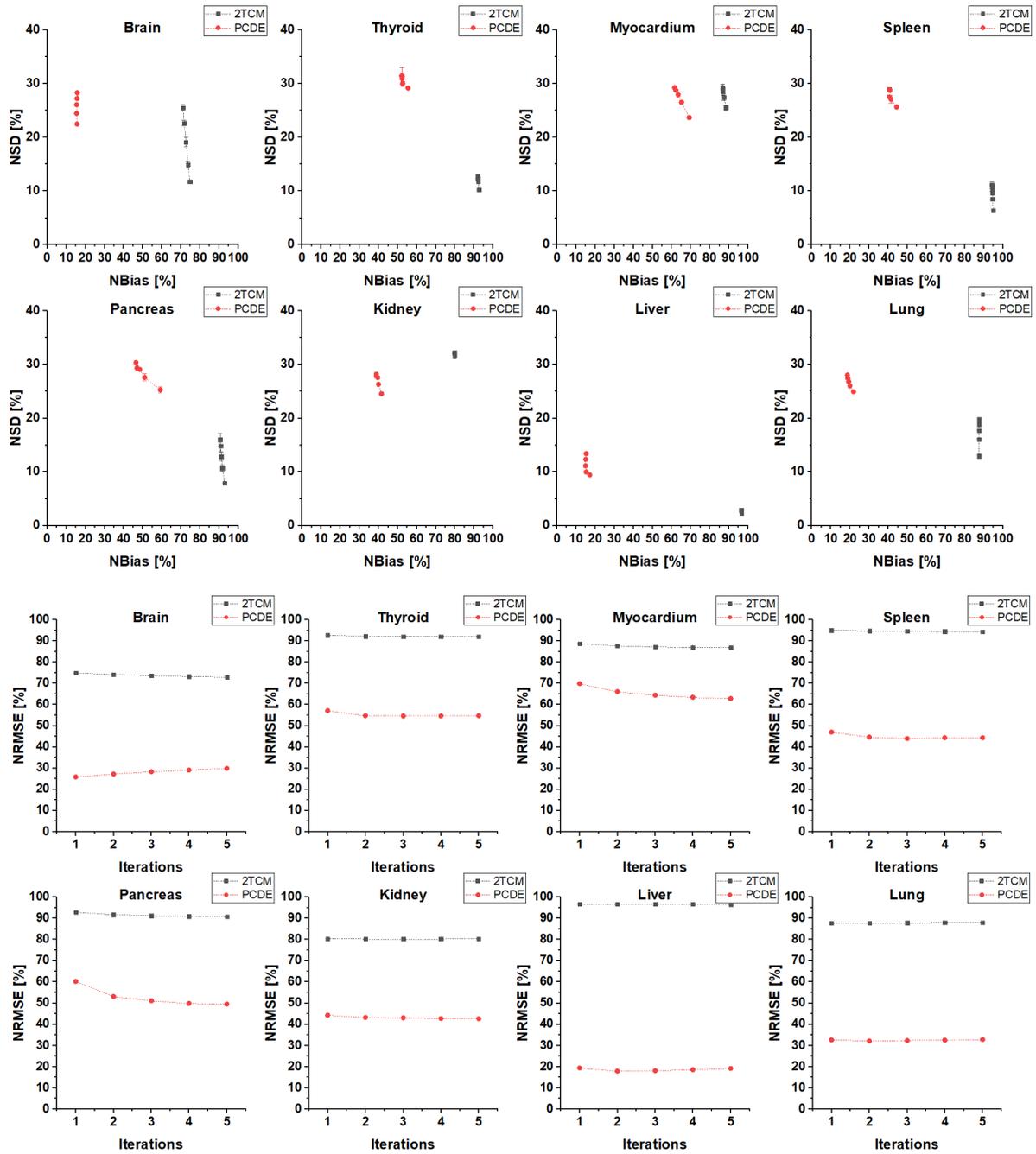

Supplemental Figure 3. ROI-based NBias-NSD trade-off curves (i.e., upper two rows) and NRMSE results with increasing OSEM iterations (i.e., lower two rows) for parametric $K_1$ images.

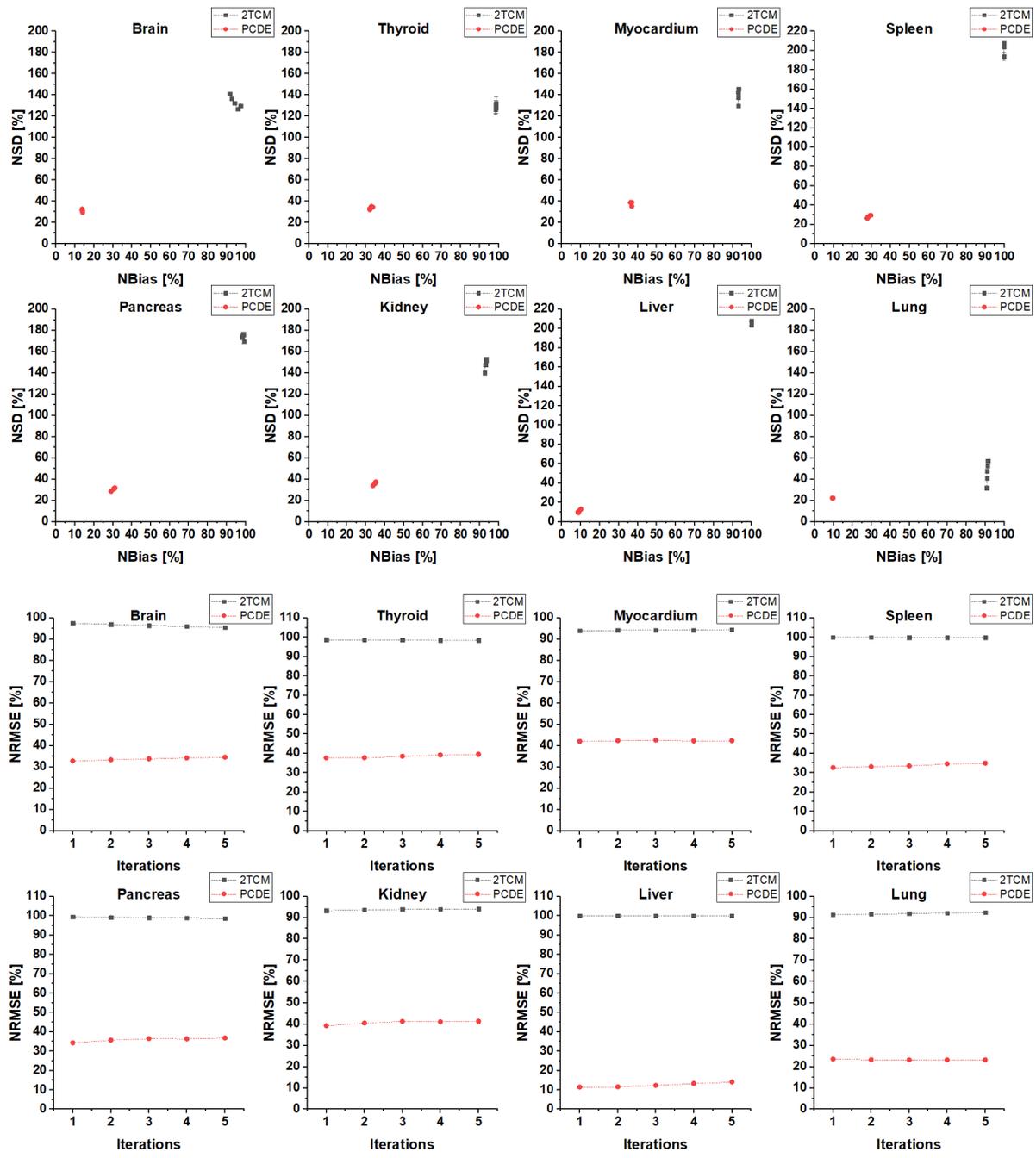

Supplemental Figure 4. ROI-based NBias-NSD trade-off curves (i.e., upper two rows) and NRMSE results with increasing OSEM iterations (i.e., lower two rows) for parametric $k_2$ images.

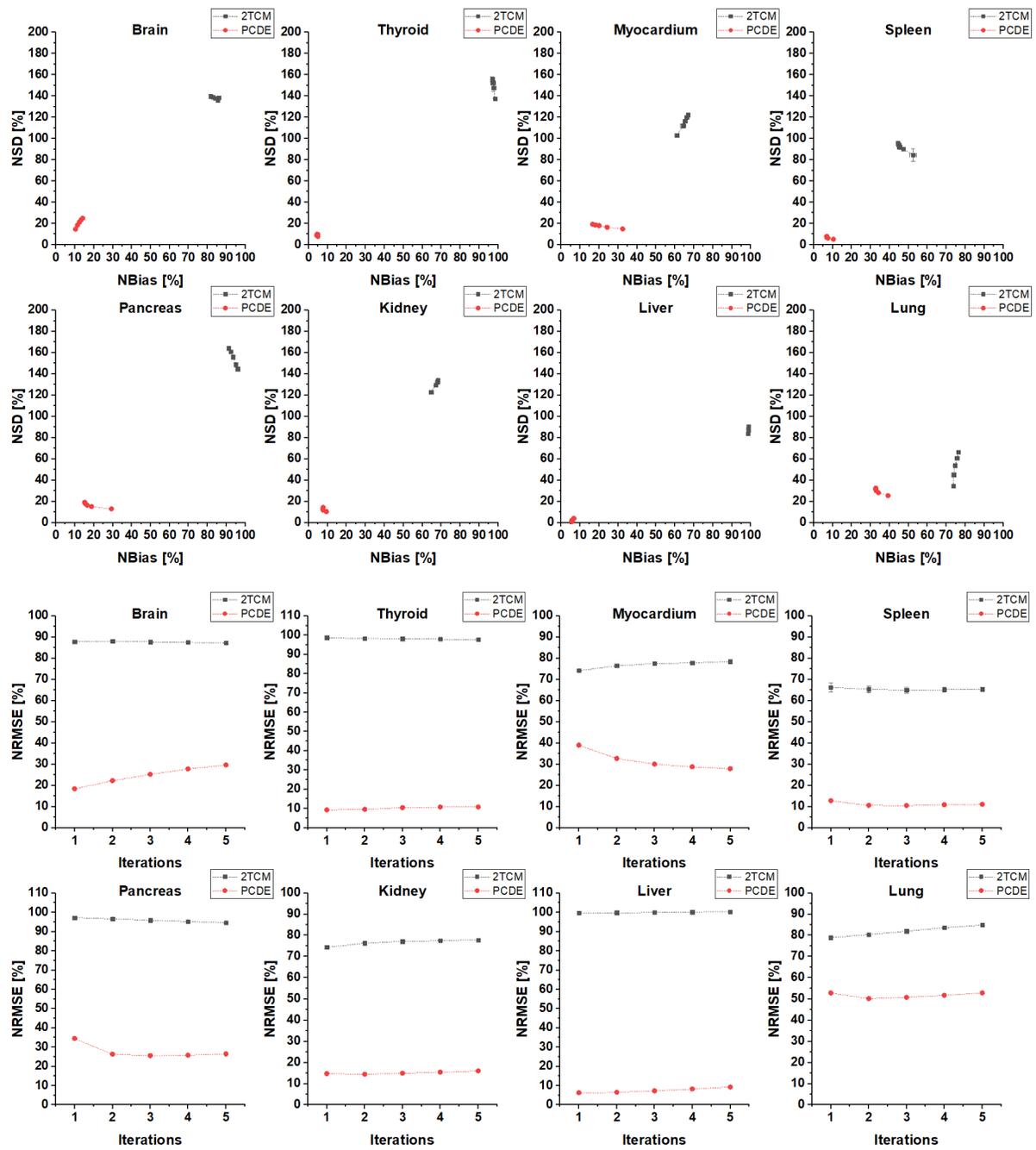

**Supplemental Figure 5.** ROI-based NBias-NSD trade-off curves (i.e., upper two rows) and NRMSE results with increasing OSEM iterations (i.e., lower two rows) for parametric $k_3$ images.

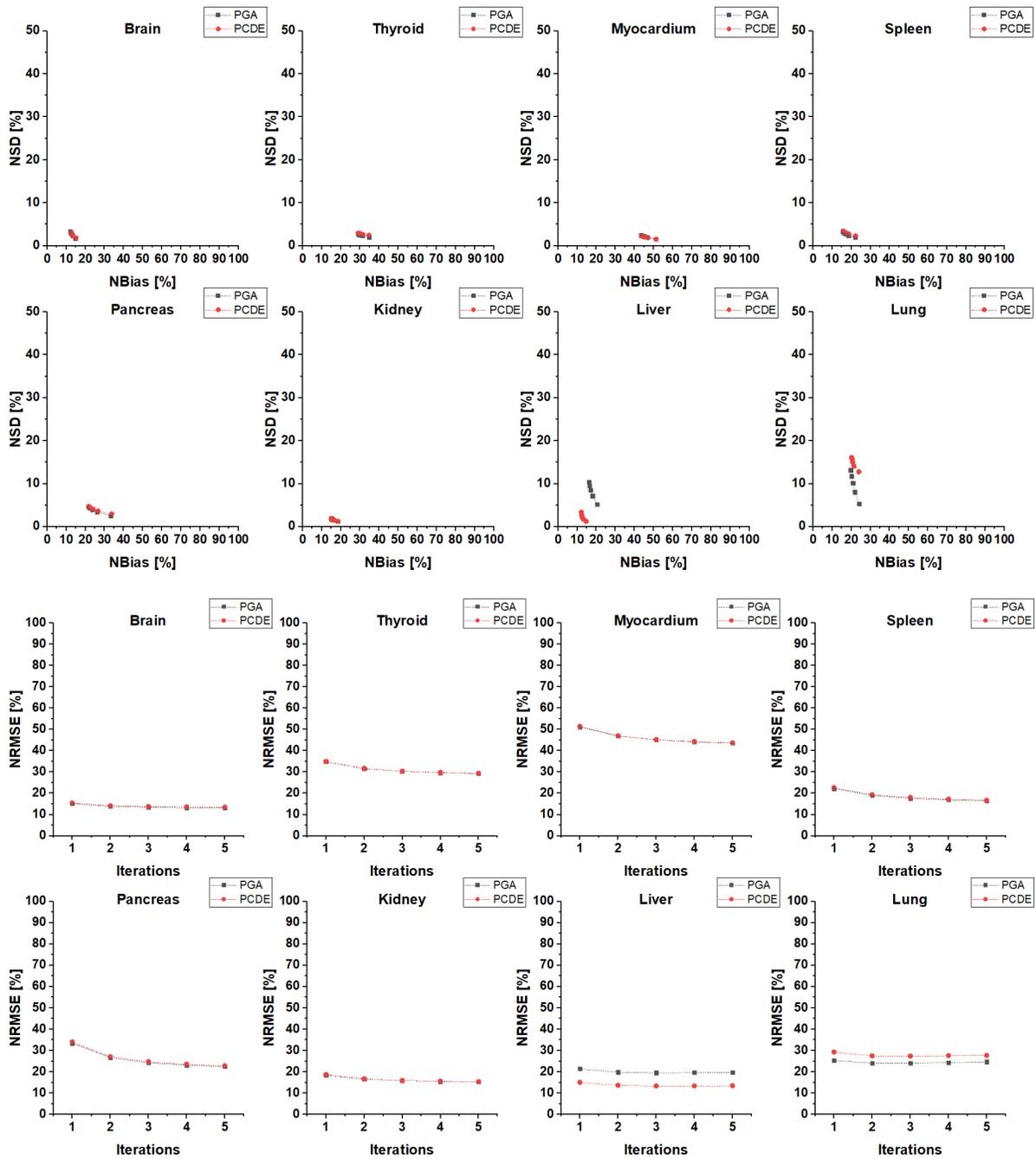

**Supplemental Figure 6.** ROI-based NBias-NSD trade-off curves (i.e., upper two rows) and NRMSE results with increasing OSEM iterations (i.e., lower two rows) for parametric $K_i$ images.

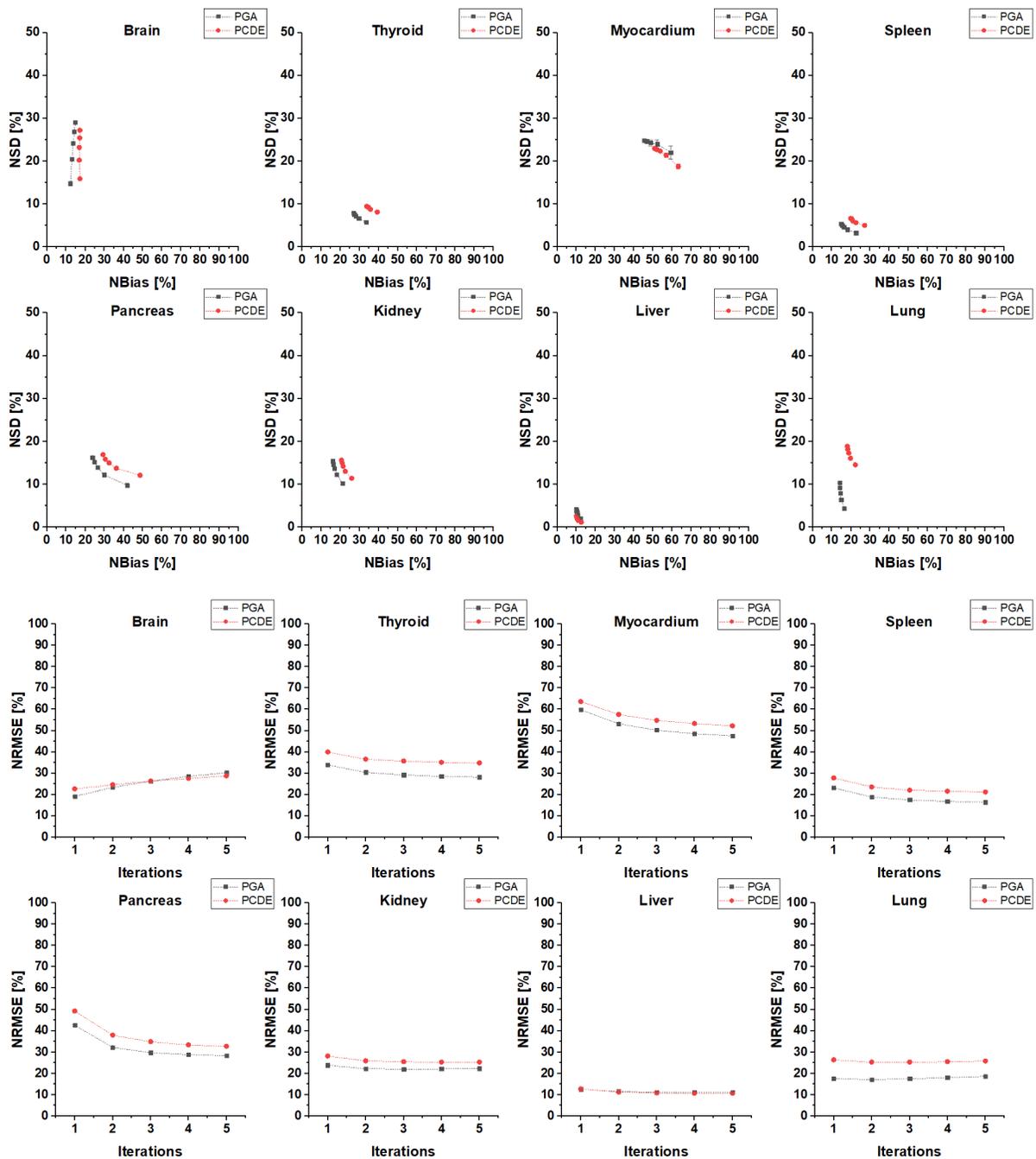

Supplemental Figure 7. ROI-based NBias-NSD trade-off curves (i.e., upper two rows) and NRMSE results with increasing OSEM iterations (i.e., lower two rows) for parametric $V_d$ images.

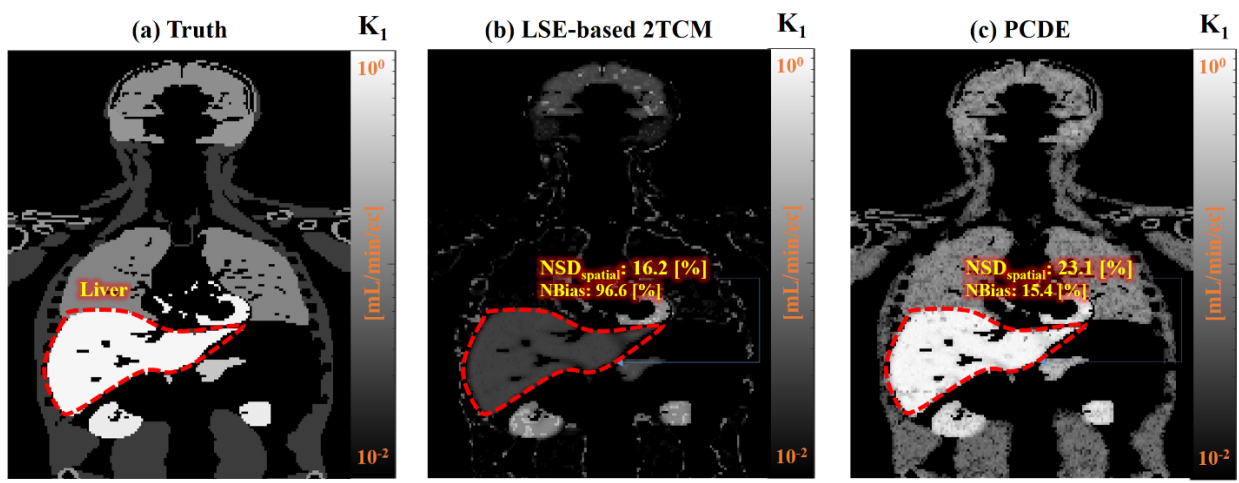

**Supplemental Figure 8.** Example of an erroneously lower level of NSD$_{spatial}$ with high bias in the LSE-based 2TCM K$_1$ image mainly due to the local minimum issue. (OSEM iterations=5, Noise realization index=1). Note that unlike NSD$_{spatial}$, each NBias values were calculated from all noise realizations.

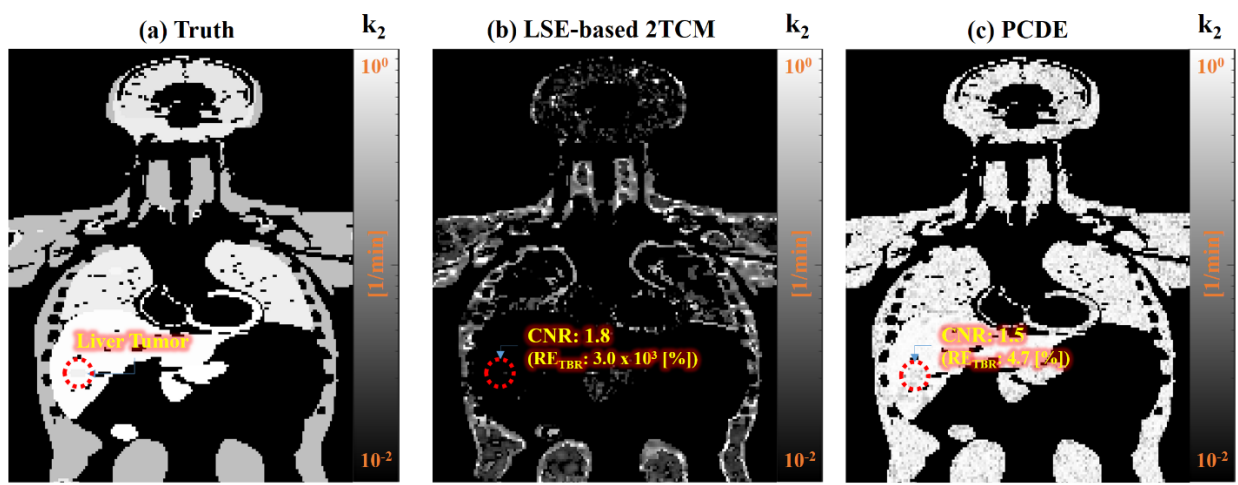

**Supplemental Figure 9.** Example of a misleading CNR and necessity of RE$_{TBR}$ for a fair comparison. (a): Ground Truth. (b): LSE-based 2TCM. (c): PCDE. (OSEM iterations=5, Noise realization index=1, Kinetic parameter: k$_2$).